\documentclass[aps,prx,reprint,floatfix]{revtex4-1}

\usepackage[pdftex]{graphicx}
\DeclareGraphicsExtensions{.pdf,.png}

\usepackage{grffile}

\usepackage{hyperref}
\hypersetup
{
pdfmenubar=true, 
pdfhighlight=/O, 
pdfpagelayout=SinglePage, 
pdffitwindow=true, 
colorlinks=true, 
linkcolor=blue, 
citecolor=blue, 
urlcolor=blue 
}

\usepackage{amsmath,amsthm, amssymb, amsfonts, amsbsy}
\usepackage{mathtools} 
\usepackage{bm}

\newcommand{\abs}[1]{\lvert #1 \rvert}
\newcommand{\norm}[1]{\left\lVert#1\right\rVert}

\usepackage[capitalise]{cleveref}
\crefformat{equation}{Eq. (#2#1#3)} 
\Crefformat{equation}{Equation (#2#1#3)} 
\crefrangeformat{equation}{Eqs. (#3#1#4--#5\crefstripprefix{#1}{#2}#6)} 
\Crefrangeformat{equation}{Equations (#3#1#4--#5\crefstripprefix{#1}{#2}#6)} 
\crefmultiformat{equation}{Eqs. (#2#1#3)}{ and~(#2#1#3)}{, (#2#1#3)}{ and~(#2#1#3)} 
\Crefmultiformat{equation}{Equations (#2#1#3)}{ and~(#2#1#3)}{, (#2#1#3)}{ and~(#2#1#3)} 

\newcommand{\dt}[1]{{d #1 \over d t }}

\newcommand{\Tr}{\mbox{Tr}}

\newcommand{\nvec}{\hat{\bf n}}

\newcommand{\ham}{{\cal H}}

\newcommand{\bfr}{{\bf r}}

\newcommand{\bfq}{{\bf q}}
\newcommand{\bfk}{{\bf k}}

\newcommand{\bfv}{{\bf v}}

\newcommand{\bmf}{{\bm f}}

\newcommand{\bmv}{{\bm v}}
\newcommand{\bmk}{{\bm k}}
\newcommand{\bmQ}{{\bm Q}}
\newcommand{\bmS}{{\bm S}}
\newcommand{\bmW}{{\bm W}}
\newcommand{\bmM}{{\bm M}}

\newcommand{\bsigma}{{\bm \sigma}}
\newcommand{\balpha}{{\bm \alpha}}

\newcommand{\bOmega}{{\bm \Omega}}
\newcommand{\caC}{{\cal C}}

\newcommand{\aver}[1]{\left\langle {#1}\right\rangle}

\usepackage{graphicx}
\usepackage{color}

\newcommand{\ds}{\displaystyle}
\newcommand{\beq}{\begin{equation}}
\newcommand{\eeq}{\end{equation}}
\newcommand{\beqa}{\begin{eqnarray}}
\newcommand{\eeqa}{\end{eqnarray}}
\newcommand{\bem}{\begin{math}}
\newcommand{\eem}{\end{math}}
\newcommand{\rar}{{\rightarrow}}
\newcommand{\uar}{{\uparrow}}
\newcommand{\dar}{{\downarrow}}

\newcommand{\bfD}{{\bf D}}

\newcommand{\vX}{{\vec X}}
\newcommand{\vnabla}{{\vec \nabla}}
\newcommand{\vv}{{\vec v}}
\newcommand{\vxi}{{\vec \xi}}
\newcommand{\vx}{{\vec x}}
\newcommand{\vy}{{\vec y}}

\def\strutdepth{\dp\strutbox}
\def\nw#1{\strut\vadjust{\kern-\strutdepth\vtop to0pt{\vss\hbox to\hsize
{\hskip\hsize\hskip5pt$\leftarrow$\hss\strut}}}{\em #1}}
\usepackage{multirow}
\def\tl#1{\textcolor{black}{#1}}

\def\tbl#1{\textcolor{black}{#1}}

\begin{document}

\title{Steady-state distributions and non-steady dynamics in non-equilibrium systems}

\author{Tanniemola B. Liverpool}
\email[]{t.liverpool@bristol.ac.uk}
\affiliation{School of Mathematics, University of Bristol - Bristol BS8 1TW, UK}

\date{\today}

\begin{abstract}
\tbl{We search for steady states in a class of} fluctuating and driven  physical 
systems that exhibit sustained currents. 
We \tbl{find 
that the physical concept of a steady state, well known for systems at equilbrium,  must} be generalised to describe \tbl{such systems. 
In these, the generalisation of 
a 
steady state 
is
} associated with a stationary probability density of micro-states {\em and} a deterministic dynamical system whose trajectories the system follows on average. 
These trajectories are a manifestation of 
non-stationary macroscopic currents observed in these systems. 
We determine precise conditions for the steady state to exist as well as the requirements  for it to be stable.
We illustrate this with some examples.
\end{abstract}


\maketitle

\section{Introduction}

The study and classification of 
non-equilibrium systems remains one of the major open problems in statistical physics~\cite{Born1946,DeGroot1984book,Jarzynski1997a,Fisher1998a,Prahofer2000,
Sasamoto2010,Seifert2012}.
A large class of non-equilibrium systems are {\em driven} systems whose behaviour is characterised by the presence of sustained non-zero currents. Unlike undriven systems whose dynamics 
is 
the 
relaxation towards equilibrium where all currents are zero, such  systems show complex dynamics :
oscillations, dynamic order-disorder transitions, pattern formation and phase separation \cite{Pedley1990,Vicsek1995,Toner1998,Dombrowski2004a,Hatwalne04,Golestanian2005,Howse2007,Sanchez2012, Deseigne2010,Jiang2010,Schaller2010,Palacci2013,Bricard2013,Cates2012,Marchetti2013,Ginot2015}.  They are also thought to be the framework for 
new theories of ``
active or 
living matter '' ~\cite{Mackintosh2008,Henkes2011a,Buttinoni2013,Attanasi2014,Solon2015a,Giomi2011,Peruani2012,Wysocki2014,Takatori2014,Speck2014,Yang2014e,Redner2016} required to describe biological systems~\cite{Alberts2007book,Camalet2000,Gardel2004,Bursac2005,Mizuno2007,Hong2007,Trepat2009,Cates2010,Polin2009,Sokolov2009,Thutupalli2011,Brugues2012,Schwarz2013,Battle2016,Saw2017}.

 
The Gibbs-Boltzmann distribution of equilibrium statistical mechanics describes the probability density of states $\rho$  of a macroscopic system 
at \tbl{equilibrium with} a fixed temperature, $T$ 
as a function of the total energy $\ham$ of the system in that state, $\rho \propto e^{- \ham/T}$ and forms the starting point for studies of many interacting particle systems at equilibrium~\cite{ChaikinLubensky95}. 
The average (statistical) properties of equilibrium systems can thus all be expressed as integrals (or sums) over this distribution though the evaluation of these integrals are in general difficult to perform~\cite{Onsager1944}. 
An important constraint on any dynamical model of a system evolving towards equilibrium is the requirement that the Gibbs-Boltzmann distribution is a {\em stable}  steady state, i.e. that the probability density of states of the system evolves towards it for long times~\cite{vankampen,Bakry1984,Bakry1994book,Markowich2000,Guionnet2002} \tbl{and once there, stays there}. 
In contrast, the situation for non-equilibrium systems  is much less clear.
Non-equilibrium systems remain largely not understood except for some special cases~\cite{Born1946,DeGroot1984book,Jarzynski1997a,Fisher1998a,Prahofer2000,
Sasamoto2010,Seifert2012}.  

\tl{In this article \tbl{our goal is to find out if (and under what conditions) 
steady states exist for some classical non-equilibrium systems. We obtain a stationary probability density of states, analogous to the Gibbs-Boltzmann distribution 
but find that this is insufficient to completely describe their behaviour. We find that the `steady-states' of these} non-equilibrium systems have an additional physical property that must be added to the \tbl{characterisation of the `steady state'} 
in order for this analogy to be made.} 
%
We \tl{show}
that for \tbl{this} wide class of driven classical systems, one can indeed find probabilistic steady states in the sense that 
the probability density of microstates achieves a steady state. However unlike equilibrium steady states,  where the system is on average stationary and fluctuates around the minimum of free energy these 
steady states are characterised by the system 
fluctuating around {\em typical trajectories} which keep the distribution constant. We show how to explicitly calculate these trajectories 
and identify conditions for the steady state to be stable.
\tl{The steady states of such systems 
are thus characterised by two linked 
mathematical 
objects, a steady state probability distribution of microstates {\em and} a {\em deterministic} dynamical system whose trajectories the system follows on average.} 
We call them {\em generalised steady states}.

Therefore the average (statistical) properties of these non-equilibrium systems can thus be expressed in terms of integrals over this distribution 
and because of the average deterministic dynamical system are in general not stationary but varying in time. 
\tl{Whilst some of our results are rigorous, our interest here is in concrete physical realisations of these steady states in experimentally feasible systems.
We study several examples of driven fluctuating systems characterised by these {generalised} steady states:  a driven oscillator, a model of a chemical reacting system and a thin film of active nematic in a disordered flowing state sometimes referred to as `active turbulence'~\cite{Sanchez2012,AditiSimha2002} .
}

\section{Non-equilibrium systems 
}
 
Non-equilibrium systems 
are defined dynamically, i.e. by a set of dynamic rules encoding their evolution.
We thus consider systems with $N>1$ ``microscopic" degrees of freedom $\vx(t) = \left(x_1(t), \ldots, x_N(t) \right)$ generically undergoing 
dynamics (a sum of deterministic and fluctuating parts) given by the Langevin equation,
\beq
{d \over dt} \vx = - {\bf D} \cdot \vec\nabla {\ham} (\vx)  + \vec w(\vx)  + \vec \xi (t)
\eeq
where $\ham (\vx)$ is  a scalar function of $\vx$, $\bf D$ is a mobility matrix and 
the gradient operator, $\ds \vnabla=({\partial \over \partial x_1}, \cdots, {\partial \over \partial x_N}) \equiv (\nabla_1, \ldots, \nabla_N)$.
 The fluctuations, $\vec \xi = (\xi_1,\ldots,\xi_N)$ are white with zero mean and autocorrelation function: 
 \beq
\aver{\xi_j (t)} =0 \; ; \quad \aver{\xi_i (t) \xi_j (t')} = 2 \theta D_{ij} \delta(t-t') \; , \; \theta > 0  \; .
 \eeq
 To be concrete we consider diagonal mobility matrices~~\footnote{The proof generalizes to non-diagonal diffusion constants in a straightforward way} of the form $D_{ij}=D_i \delta_{ij}$ where $D_i>0$ are independent of $\vx$ and  $\delta_{ij}$ is the Kronecker delta~\cite{TBL_unpub}.
 We can w.l.g rewrite $\vec w(\vx)= {\bf D} \cdot \vec v(\vx)$ where the differentiable vector-valued function $\vec v (\vx)=(v_1(\vx),\ldots, v_N(\vx))$ cannot be written as the derivative of a scalar function.
 This implies the microscopic breaking of  ``detailed balance''~\cite{vankampen}.
 Equations like this emerge in 
 many 
 models for 
 slow dynamics of driven physical systems~\cite{Doi1986book,ChaikinLubensky95}.

 The Langevin equation is equivalent to a Fokker-Planck equation~\cite{Risken} for the probability density, $P(\vx,t)$:
 \beq
 \partial_t P = \sum_{i=1}^N  \nabla_i D_i  \left( \theta \nabla_i  P + P ( \nabla_i \ham - v_i) \right) \; .
 \label{eq:F-P}
 \eeq
We assume $P$ is well behaved, i.e. $P \, , \, \nabla P \rar 0$ as $| \vx | \rar \infty$. We find a steady state probability density, by requiring RHS of eqn. (\ref{eq:F-P}) to vanish. 
However even 
more important 
is determining {if it is stable}, i.e. that the system moves towards it and remains there. This is our goal.
We now state our main result more formally.

 \paragraph*{Definition:}
 
\tl{ 
It is useful to define a function $h(\vx)$ as follows: 
 the system has a 
 steady state probability density $P_{ss}=\rho (\vx) = \frac1Z e^{- h(\vx)}$ if a function $h(\vx)$ can be found that 
  satisfies
 \beq
  \sum_{i=1}^N D_i  L_i (h) =0 \label{eq:ss-cond}
 \eeq
where $L_i (h) = \theta (\nabla_i h)^2 + \nabla_i^2 \ham +
 \nabla_i h \left( 
 v_i-\nabla_i \ham  \right)- \theta \nabla_i^2 h   - \nabla_i v_i  $
and the normalisation $Z=\int d^N x \, e^{-h(\vec x)}$ is chosen so that $\int d^N x \rho (\vx)=1$. }
%

 \paragraph*{Theorem 1a:}
 
 \tl{
 For functions $h(\vx)$ which satisfy eqn. (\ref{eq:ss-cond}), $\rho$ remains constant on the trajectories: $\vx (t) = \vX (t)$, 
\beq
{d X_i \over dt} =  V_i (\vX)  \quad , \quad V_i = D_i \left( v_i - \nabla_i \ham + \theta \nabla_i h \right)
\eeq
The set $\{ \rho, \vec V \}$  characterise a generalized steady state.}

\tl{A useful decomposition of the equation (\ref{eq:ss-cond}) are 
solutions, $h(\vx)$  which satisfy both the following conditions :
\beq
 \sum_{i=1}^N  \nabla_i V_i  = - \caC \quad , \quad \sum_{i=1}^N  V_i \nabla_i h = \caC \quad ,
\eeq
and we base our subsequent analysis on this observation.}

\paragraph*{Theorem 1b:}  
\tl{
 If $\caC(\vx)  \ge 0$, $\forall \;  \vx $ then the generalized steady state is stable and the system will always evolve towards it for any arbitrary initial condition.
 If not, i.e. if $\caC < 0$ for some values of $\vx$ then the situation is inconclusive, the steady state may be generically unstable or the stability of the steady state may depend on initial conditions and the values of parameters.}
 
 \tl{To obtain more information about how quickly the system relaxes to the generalised steady state, 
 it is helpful to decompose the set of stable scenarios into two classes.
 \paragraph*{Theorem 2a:}
 If $\caC=0$, then 
 the system relaxes exponentially fast to the steady state if  all the eigenvalues of the Hessian matrix 
$\nabla_i \nabla_j h$ are all positive.
\beq
\vnabla\vnabla h > 0 \quad.
\eeq
 Note that ``equilibrium'' systems with $v_i(\vx)=0$ have $h= \ham (\vx)/\theta$ (the Boltzmann distribution at temperature $\theta$),  $V_i=0$ and hence $\caC=0$. 
 }

 \paragraph*{Theorem 2b:}
 
 \tl{
 If $\ds \caC>0$ and $\ds \lim_{| \vx | \rar 0} \caC >0$ then the system relaxes exponentially fast to the stationary state, irrespective of the form of $h(\vx)$. Such systems I denote as {\em super stable}.
 }
 
 \tl{We note that while systems satisfying these conditions {\em will always} relax exponentially fast, 
it is also quite possible that systems which {\em do not} satisfy them might also relax exponentially fast under certain conditions, and that the bounds we have obtained can be improved to include a wider class of systems. 
Furthermore,  it is important to note that when $\caC=0$ and $\vnabla\vnabla h  \not > 0$, then the system {\bf is} still stable, just that it could possibly relax very slowly to the generalised steady state.
We also point out out that these results are valid for arbitrarily large noise amplitude, $\theta$.}
 
\tl{  
By obtaining $V_i(h) \ne 0$, we have explicitly calculated the macroscopic current.
When the amplitude of the noise, $\theta=0$, the typical trajectories are those of the deterministic equation. As other trajectories do not keep the probability density constant, typical trajectories act as attractors. Finally, we note that in one dimension, $N=1$, the only possible steady state dynamical system is one with $\rho_{} V$ constant.
}

  We emphasize that the statement that $h(\vx)$ which satisfies eqn. (\ref{eq:ss-cond}) determines a probability density that is stationary is by itself not particularly helpful or new~\cite{Gardiner1996}. 
 This is because in practice  the nonlinear steady state equation will yield several (in general, approximate) solutions for $h$, and it will not be obvious even if any of them is stable, i.e. an  attractor for the dynamics. The Fokker-Planck equation can be written as $\ds \partial_t P + \sum_{i=1}^N \nabla_i J_i = 0 $ where $J_i$ is a probability current; the stationary condition, eqn. (\ref{eq:ss-cond}) is simply the statement that 
 the steady-state is associated with a divergenceless current, $\ds  \sum_{i=1}^N \nabla_i J_i=0 \; $~\cite{Gardiner1996}.  This gives rise to a complicated non-linear partial differential equation for $h$ which may have no, or more than one, solution.
Furthermore since it is determined by a nonlinear PDE, one must have a method to calculate it (even approximately). Hence identifying conditions that make $\rho\propto e^{-h}$ stable and outlining a systematic way to obtain $h$ is the main result of the theorem above and the subject of this article. 
We will be particularly interested in situations where  $h=\ham/\theta + \epsilon$ with $ \epsilon < \ham/\theta$; this is the case for many examples of active matter~\cite{Marchetti2013}.

We now outline a proof below.
 
 \paragraph*{Proof:}
 
 \tl{
 That $h$ determines a stationary probability density and remains constant on typical trajectories follows trivially by substitution.} To show that it determines a stable (in the probabilistic sense) probability density,
we show that if we start with an arbitrary density $P(\vx,t) = \rho (\vx)  \pi (\vx,t)$, with the conditions above on $h$, $\pi \rar 1$  and $d \vx /dt \rar \vec V$ exponentially fast.
 Substituting this density into eqn. (\ref{eq:F-P}) we get a modified backward Kolmogorov equation for $\pi$:
 \beq
 \partial_t \pi = {\cal L} \pi \quad , \quad {\cal L} = \sum_i D_i \left\{  \theta \nabla_i^2  +  W_i  \nabla_i  \right\}
 \eeq
 where
 $W_i =  \left( \nabla_i \ham  -v_i  - 2 \theta  \nabla_i h\right) $.
 Moving along trajectories , $ \dot X_i = V_i$, $\pi$ evolves according to the comoving time derivative
 \beq
 {d \pi \over d t} = \partial_t \pi + \sum_i V_i \nabla_i \pi = {\cal L}'\pi
 \eeq
 where ${\cal L}' = \sum_i \theta D_i \{ \nabla_i^2 - \nabla_i h \nabla_i\}$.

We sum over all trajectories by integrating over all possible deviations from the typical trajectories.
Defining the inner product 
 \(
 \aver{f,g}_\rho \equiv \int d^N y \rho (\vx) f (\vx) g(\vx) \, , 
 \) where $ \vx  = \vX  + \vy$, 
 and the norm $\norm{A}_\rho^2 = \aver{A,A}_\rho$,
 then for any $C^2$ function $f(\vx)$ using integration by parts, it is easy to show that
 \beq
 \aver{{\cal L}' f, f}_\rho = 
 - \theta  \sum_i D_i || \nabla_i f ||_\rho^2 \; .
 \eeq
 \tl{ To show that motion along these trajectories is stable to noise we can look for the dynamics of the deviation of the probability density from the steady state, 
 \(\ds
 \norm{{P - \rho \over \rho}}_\rho =\left[ \int d^N y \, \rho \,  (\pi-1)^2\right]^{1/2} = \norm{\pi-1}_\rho
 \)
 given by 
 \beqa
 {d \over dt}  \norm{\pi-1}_\rho^2 &=& 2 \aver{ {d \pi \over dt}, \pi-1}_\rho  - \int d^N y \, \caC(\vx)  \rho \,  (\pi-1)^2 \nonumber \\  &=& 2 \aver{{\cal L}' (\pi-1),\pi-1}_\rho  - \aver{ \caC , (\pi-1)^2}_\rho \nonumber \\
 &=& - 2 \sum_{i=1}\theta  D_i \norm{\nabla_i (\pi-1)}_\rho^2  - \aver{ \caC , (\pi-1)^2}_\rho  \; . \nonumber 
 \eeqa
 Hence,  if $\caC(\vx) = \vec V \cdot \vnabla h   \ge 0$, 
 \beq
 {d \over dt}  \norm{\pi-1}_\rho^2  \le  0 \; ,
 \eeq 
  where $V_i = D_i \left(v_i -\nabla_i \ham + \theta \nabla_i h \right)$
 \footnote{
 This condition is sufficient for stability but is not necessary, e.g., 
 one may obtain a weaker condition by noting that $ \int d^N y \, \caC  \rho \,  (\pi-1)^2 > \inf_{\vx \in {\Bbb R}^N} \left .  \, \left( \pi ( \vx )-1\right)^2 \right . \int d^N y \, \caC  \rho \, $ so stability is achieved as long as $\int d^N y \, \caC(\vx)  \rho (\vx)  \ge 0$ even if $\caC(\vx) < 0$ for some $\vx$.}.
 }
 \tl{This proves that $\norm{\pi-1}_\rho$ always decreases with time if $\caC \ge 0$. However one would also like to know how quickly the system relaxes to the steady-state. 
 In what follows we set all the $D_i=1$ to simplify formulas.}
 
 \tl{First we consider the case $\caC=0$.
 For this we use  Bakry-\'Emery inequality~\cite{Bakry1984,Bakry1994book,Markowich2000,Guionnet2002}.
 The inequality is obtained in this setting by  taking the comoving time derivative of $ \norm{\nabla_i (\pi-1)}_\rho^2$, integrating by parts:
 \beqa
 {d \over d t}  
\norm{\nabla_i (\pi-1)}_\rho^2  &\leq& -\sum_{ j } \frac{\theta 
}2 \int_\vy \rho {\nabla_i \pi } {\nabla_j \pi } 
 \nabla_i \nabla_j h \nonumber \\
 &\le& - 
 \frac{\theta 
 \lambda_0}{2} \norm{\nabla_i (\pi-1)}_\rho^2
 \eeqa
 where $\lambda_0>0$ is the smallest eigenvalue of the Hessian matrix $\vnabla\vnabla h$. Hence once $\vnabla\vnabla h >0$, $ \norm{\nabla_i (\pi-1)}_\rho^2$ and consequently $\norm{\pi-1}_\rho^2$ relax exponentially fast to zero on a timescale of order $(\theta  \lambda_0)^{-1}$.
 }
 
 \tl{Next we consider the case $\caC >0$ and $\lim_{\vx \rar 0} \caC(\vx) = \caC_0  >0$. Here, 
 \beqa
 {d \over dt}  \norm{\pi-1}_\rho^2 & \le & - \caC_0  \norm{\pi-1}_\rho^2 \; ,
 \eeqa
and $\norm{\pi-1}_\rho^2$ relaxes exponentially fast to zero on a timescale of order $(\caC_0)^{-1}$, irrespective of the form of $h$ as long as $\caC,\caC_0 >0$.
 %
Clearly if there are several (possibly approximate) values for $h$, this provides a way to rank them.}

 The proof above relied on being in a finite-dimensional vector space $\vx \in {\Bbb R}^N$, hence
 these results can  be 
 generalized to {\em regularised} stochastic field $\bmf(\bfr,t)$ dynamics 
 %
 where 
 \bem \; \bfr \in L^d  
 \quad
 \eem 
 in the following sense.
 We define an expansion (e.g Fourier) in a set of orthonormal basis functions , 
 \(\ds
 \bmf(\bfr,t) = \sum _{\bfq}^{} \bmf_{\bfq}(t)  \Psi_\bfq(\bfr)\; , \; \int_\bfr \Psi_\bfq^*(\bfr) \Psi_{\bfq'} (\bfr)= \delta_{\bfq' \bfq}  \)  (for Fourier series,  \(\Psi_\bfq=e^{i \bfq \cdot \bfr} \; ; \; {L \over \pi} \bfq \in {\Bbb Z}^d 
 \))
 which can be regularized by restricting the number of modes to a finite number,
 \(\ds
 \bmf_\Lambda(\bfr,t) = \sum _{\bfq=\bfq_{min}}^{\bfq_{max}} \bmf_{\bfq}(t) \Psi_\bfq(\bfr) \) .
In the Fourier expansion,  $|\bfq_{min}| \sim \pi/L$ and  $|\bfq_{max}| \sim \pi/a$ where $a$ is a short-distance lengthscale.  
The restricted  modes $\{\bmf_\bfq\}$ are a finite vector space with $N \gg 1$ 
 which satisfy the theorem above.
%
%

Now we illustrate the theorem with some examples for which we calculate macroscopic currents. 
It turns out that many 
examples of driven active systems 
have 
$\caC=0$.

\section{Examples}

\subsection{The noisy Hopf oscillator}

\tl{ The study of the effects of fluctuations on the normal form of an oscillator that can go through a Hopf bifurcation provides a relatively simple non-trivial two dimensional system where one can study the implications of our main result.
The degrees of freedom $\vx (t) =(x_1,x_2)$  have equation of motion
\beq
\dt \vx = A \vx - B |\vx|^2 \vx + \bOmega \cdot \vx + \vec \xi (t) \quad , 
\eeq
where
\bem
\ds \bOmega = \left(  \begin{array}{cc} 0 & \Omega \\ - \Omega & 0 \end{array} \right) \, , \quad
\eem
and $A,B>0,\Omega$ are constants.
The noise $\vxi = (\xi_1, \xi_2)$ has zero mean and mean square fluctuations
\beq
\aver{\xi_i(t) \xi_j(t') }= 2 \theta \delta_{ij} \delta(t-t') \quad.
\eeq
This is of the form 
\beq
\dt \vx = - \vnabla \ham + \vv (\vx) + \vxi (t) \quad, 
\eeq
where $\ham = -\frac{A}{2} |\vx|^2 +\frac B4 (|\vx|^2)^2 $ and  $\vv = \bOmega \cdot \vx$ cannot be written as the gradient of a scalar function. For vanishing noise amplitude, $\theta=0$ the system is deterministic and for $A>0$ undergoes oscillations with frequency $\Omega$ and amplitude $\sqrt{A/B}$.
}

\tl{We look for solutions of the form \( \ds h = {\ham \over \theta} + \epsilon \,, \) where
\beq
  \epsilon = \frac12 \vx \cdot \bmM \cdot \vx + \frac{g}{4} | \vx |^4 + \ldots \quad , \quad \bmM = \left(  \begin{array}{cc} m_1 & m_2 \\ m_4 & m_3 \end{array} \right) \, , 
\eeq
which is reasonable if $|\vx|$ is not too large.
We allow the most general quadratic form and assume the form of the stabilising quartic term is unchanged but that its coupling constant may change.
This can be substituted into eqn.~(\ref{eq:ss-cond}) to obtain a power series which can be set to zero term by term starting with the lowest powers to obtain simultaneous nonlinear equations for the coefficients, $m_i,g$~\cite{Appendix}.
\tbl{We find one solution, $m_1=m_2+m_4=m_3=g=0$,  which when we expand 
around the critical point of the corresponding expression for $h(\vx)$,  
 has two positive eigenvalues for the matrix $\nabla_i \nabla_j h $.  We 
thus expect the system to relax exponentially fast to this distribution, i.e. $\rho \propto e^{-\ham/\theta}$, so we have shown 
 that for this system, the steady state distribution has the same form as the equilibrium one  as long as the typical value of $| \vx |  \sim \sqrt{A/B}$ is small enough~\footnote{For larger values of $| \vx |$ we must include higher order terms in the expansion.}.} The preferred trajectories are thus given by $V_i= v_i$ and correspond to oscillations with frequency $\Omega$. The effect of fluctuations is to generate a cloud of points around the deterministic limit cycle.
}

\subsection{
The noisy Brusselator:}

\begin{figure*}
	\begin{flushleft}
		(a) \hspace{0.43\linewidth} (b)
	\end{flushleft}
	\vspace{-1em}
	\centering
	\includegraphics[width=0.49\linewidth]{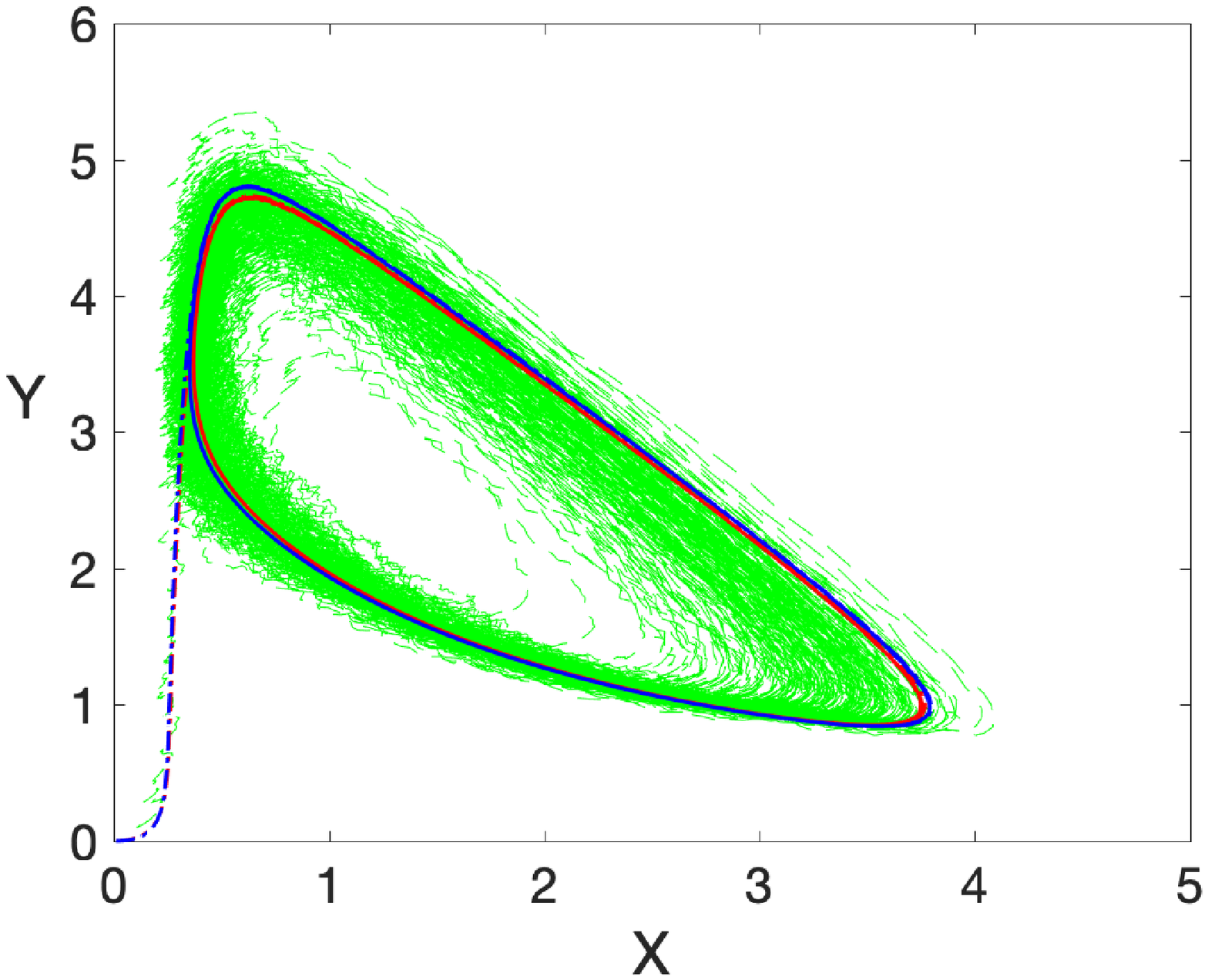}
	\includegraphics[width=0.49\linewidth]{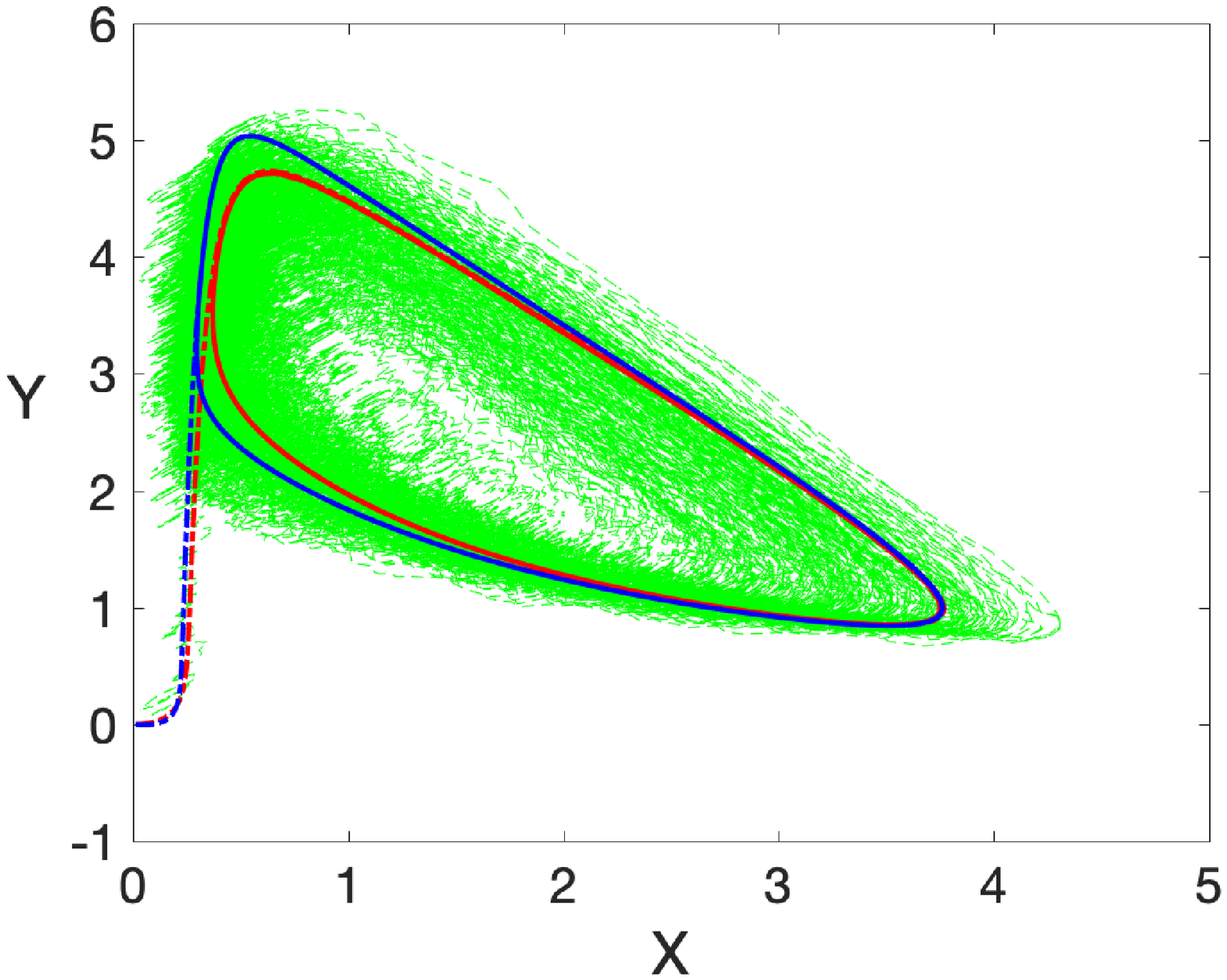}
\caption{	\tl{Phase portrait of the noisy Brusselator  with (a)  $\mu=1,\lambda=3,\theta=\frac12 (0.1)^2$; (b)  $\mu=1,\lambda=3,\theta=\frac12 (0.2)^2$. The green curve is a stochastic trajectory, the red curve is the trajectory of the deterministic Brusselator and the blue curve the typical trajectory of the noisy system.}
		\label{fig:brusselator}
	}
\end{figure*}

A more complicated example is provided by the effect of fluctuations on the dynamics of the Brusselator.
The Brusselator is a simple two dimensional dynamical system that shows oscillatory behaviour~\cite{Strogatz94}. We now use the results above to study the effects of fluctuations on this system.
We consider equations for species $x,y$ 
\beqa
\dt{x} &=& \mu + x^2 y  - \lambda x - x  + \xi_1 (t) \nonumber \\
\dt{y} &=& \lambda x - x^2 y + \xi_2 (t) \label{eq:noisy-bru}
\eeqa
with $\aver{\xi_i}=0$ and \(
\aver{\xi_i(t) \xi_j(t') }= 2 \theta \delta_{ij} \delta(t-t')
\).
In the absence of fluctuations, $\theta=0$ the system has a fixed point at $x^* = \mu, y^*=\lambda/ \mu$ which becomes unstable to oscillations for 
\(
\lambda  > 1+\mu^2 \; .
\)
We now systematically construct an expression for the steady-state density $\rho=\frac1Ze^{-h}$ when $\theta \ne 0$.
In this system $\ham = 0$ and $\vec v = (\mu+x^2y - (\lambda+1)x, \lambda x -x^2y)$. 
As long as $x,y$ are not too large, we may look for a power series expansion for $h(x,y)
$ 
\tl{and keep terms up to a particular order, e.g. 4th order : 
\(
h = a_1 x + \frac{1}{2}  a_2 x^2 + \frac{1}{3}  a_3 x^3 + \frac14 a_4 x^4 + b_0 y + b_1 xy + \frac{1}{2}  b_2 x^2 y + \frac13 b_3 x^3 y + \frac12 c_0  y^2 + \frac12 c_1  x y^2  +  \frac14 c_2 x^2 y^2 + \frac13 d_0 y^3 + \frac13 d_1 x y^3 + \frac14 e_0 y^4 \ldots
\)
This can be substituted into eqn.~(\ref{eq:ss-cond}) to obtain a power series which can be set to zero term by term starting with the lowest powers to obtain simultaneous nonlinear equations for the coefficients, $a_i,b_i,c_i,d_i,e_i$~\cite{Appendix}.}
%
Once we have the steady state we can obtain the typical trajectories.
These are 
given by $\ds \dt{\vec X} = \vec V = (\mu+x^2y - (\lambda+1)x + \theta \partial_x h , \lambda x -x^2 y + \theta \partial_y h)$.
The solution is most illustrative if we consider particular parameters. 
\tl{In Fig. \ref {fig:brusselator} (a,b), we plot a single trajectory from a solution of the stochastic differential equation (SDE) for the noisy Brusselator, eqn. (\ref{eq:noisy-bru}) with $\mu=1,\lambda=3$ for each of the two values of $\theta=\frac12 (0.2)^2, \theta=\frac12 (0.1)^2$ above 
plus  a trajectory of the deterministic Brusselator ($\theta=0$) and the typical trajectory of the noisy system. }   
We see that the effect of  the noise is to  shift the typical trajectory from the deterministic limit cycle to a new limit cycle as well as of course
generating a cloud of state points around the new limit cycle. 

\subsection{
Fluctuating d=2  active nematic:}

Active matter consists of interacting self-driven particles that individually consume energy and collectively generate motion and mechanical stresses in the bulk \cite{Toner2005,Ramaswamy2010,Marchetti2013,Saintillan2013,Prost2015}.
Due to the orientable nature of their constituents, active suspensions can exhibit liquid crystalline order and have been modeled as active liquid crystals (LCs) \cite{Ramaswamy2010,Marchetti2013,Prost2015}. 
An astonishing property of active LCs is their ability to spontaneously flow in the absence of any mechanical forcing \cite{AditiSimha2002,kruse04,Voituriez2005,Marenduzzo2007b,Giomi2008,
Furthauer2012,Giomi2013,Thampi2013}. 
We study a 2d active nematic film in the $\mbox{Re}=0$ limit. The degrees of freedom of the system are a local nematic order parameter ${\bm Q}(\bfr,t)$, traceless symmetric $2\times 2$ matrix and local fluid velocity $\bmv(\bfr,t)$, a 2d vector. 
The equations of motion are those of nematodynamics augmented to include activity~\cite{DeGennesProst93,Olmsted1990},
\beqa
0 &=&\eta \nabla^2 v_i + \delta_{ij}^T\left[ \nabla_k \sigma_{kj} + \xi_j^v (\bfr,t)\right] \label{eq:stokes} \\
\left(\partial_t + \bmv \cdot \nabla \right) Q_{ij} &=& \Omega^v_{ij}  
+ \Omega^r_{ij}  + \xi_{ij}^Q (\bfr,t) \label{eq:Qdyn}
\eeqa
where $\delta_{ij}^T= \delta_{ij} - \partial_i\partial_j /\nabla^2$, $\nabla^2 = \sum_i \partial_i^2$, 
\beq
\Omega_{ij}^r = \frac1\gamma H_{ij} \; , \;  \Omega_{ij}^v= \lambda \norm{\bmQ} u_{ij} - \left(\omega_{ik} Q_{kj} - Q_{ik} \omega_{kj} \right) \; , \nonumber  \eeq
\( \mbox{with} \;  u_{ij}=\frac12 ( \partial_i v_j +  \partial_j v_i ) \; ,  \; \omega_{ij}=\frac12 ( \partial_i v_j -  \partial_j v_i) \; ,
\) 
and
\beqa
&& H_{ij} (\bfr,t)  = - \delta F / \delta Q_{ij} (\bfr,t) \; , \nonumber \\  \;  && F =\int_\bfr \left[-\frac{A}{2} \norm{\bmQ}^2 + \frac{B}{4} \norm{\bmQ}^4 + \frac{K}{2} \left( \partial_i Q_{jk}\right)^2 \right] \;, \nonumber  \eeqa
\beq
\sigma_{ij}=-\lambda \norm{\bmQ} H_{ij} + Q_{ik}H_{kj} -  H_{ik}Q_{kj} + \alpha Q_{ij} \; ,   \nonumber \eeq 
where  $\norm{\bmQ}^2=\Tr{\,\bmQ^2}$. 

The fluctuations have zero mean $\aver{\bm{\xi}^Q}= \bm 0,  \aver{\xi_j^v}=0$. 
$\bm{\xi}^Q$ is traceless symmetric with $\xi_{11}^Q= \xi_1^Q, \xi_{12}^Q= \xi_2^Q$, 
\[
\aver{\xi_i^Q (\bfr,t) \xi_k^Q (\bfr',t') } = {2  \theta \over \gamma}  \delta_{ik} \delta(\bfr-\bfr')\delta(t-t')
\] and 
\[
\aver{\xi_i^v (\bfr,t) \xi_k^v (\bfr',t') } = -  4 \theta \eta (\delta_{ik}\nabla^2 + \partial_i \partial_k) \delta(\bfr-\bfr')\delta(t-t')
\; .\]
%
We set $\gamma=1,\theta=1$. The active contribution to the stress $\sigma_{ij}^a=\alpha Q_{ij}$ is what makes this system non-equilibrium in the manner described above and we denote $\alpha=0$ as passive and $\alpha \ne 0$ as active~\footnote{All the parameters of the model could in principle be functions of $\alpha$. However they could then be mapped on to an effective equilibrium system. Only this term cannot be removed in such a manner.}. This system shows a generic instability of the nematic ordered state to a disordered flowing state~\cite{Voituriez2005,Ramaswamy2010,Marchetti2013,Sanchez2012}. Characterising this state quantitatively in the presence of fluctuations remains an open question which we address below.

There is only one independent component of the strain rate tensor, e.g. vorticity, $\omega (\bfr,t)=\omega_{12}$.
The linearity of the Stokes eqn. (\ref{eq:stokes}) means that $\omega$ is slaved to $\bmQ$,
so that the only truly independent fields are $Q_1(\bfr,t)=Q_{11}$, $Q_2(\bfr,t)=Q_{12}$. We consider the system in a square box of area ${\cal A}=L^2$, $\bfr=(x,y), \{0 \le x,y \le L\}$ with $Q_1=\bar Q_1=S_0$ and $Q_2=\bar Q_2=0$ on the boundary $\partial {\cal A}$ where $S_0=\sqrt{2A/B}$. We consider 
deviations of $Q_i=\bar Q_i + \delta Q_i$ around the value on the boundary and $\omega$ around a stationary fluid.  We have boundary conditions, $\delta Q_1=0, 
\nvec \cdot \nabla \delta Q_2 =0$ 
on $\partial {\cal A}$.
%
%
Taking Fourier transforms, 
\bem
 \tilde Q_1 (\bfq,t) =  
 \int_\bfr \Psi_\bfq(\bfr) \delta Q_1 (\bfr,t) \, , 
\;  \tilde Q_2 (\bfq,t)  = 
 \int_\bfr \Phi_\bfq(\bfr) \delta Q_2 (\bfr,t) \, , 
\eem
with similar expressions for $\tilde\xi^Q_i(\bfq,t), \tilde\xi^v_i(\bfq,t),\tilde \omega (\bfq,t)$,
where $\Psi_\bfq(\bfr)={\cal N} \sin (q_1 x) \sin (q_2 y) $, $\Phi_\bfq(\bfr)= {\cal N}\cos (q_1 x) \cos (q_2 y)$ with $\cal N$ chosen so that $  \frac{1}{L^2} \int_\bfr \Psi_\bfq^2 =   \frac{1}{L^2} \int_\bfr \Phi_\bfq^2 =1$ and  $\bfq=(q_1,q_2)=\frac{\pi}{L}(n,m)$ with $n,m \in {\Bbb Z}^+$~\cite{Appendix}.

\begin{figure*}
	\begin{flushleft}
		(a) \hspace{0.34\linewidth} (b) \hspace{0.26\linewidth} (c)
	\end{flushleft}
	\vspace{-1.2em}
	\centering
	\includegraphics[width=0.327\linewidth]{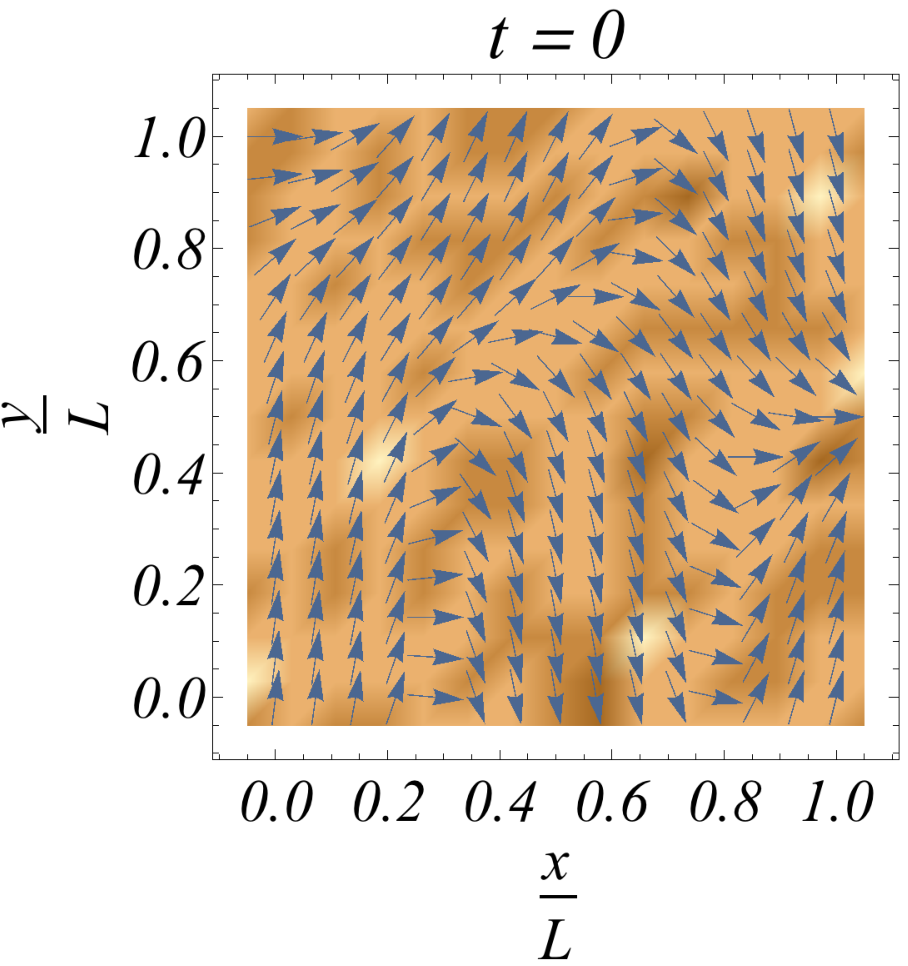}
	\includegraphics[width=0.28\linewidth]{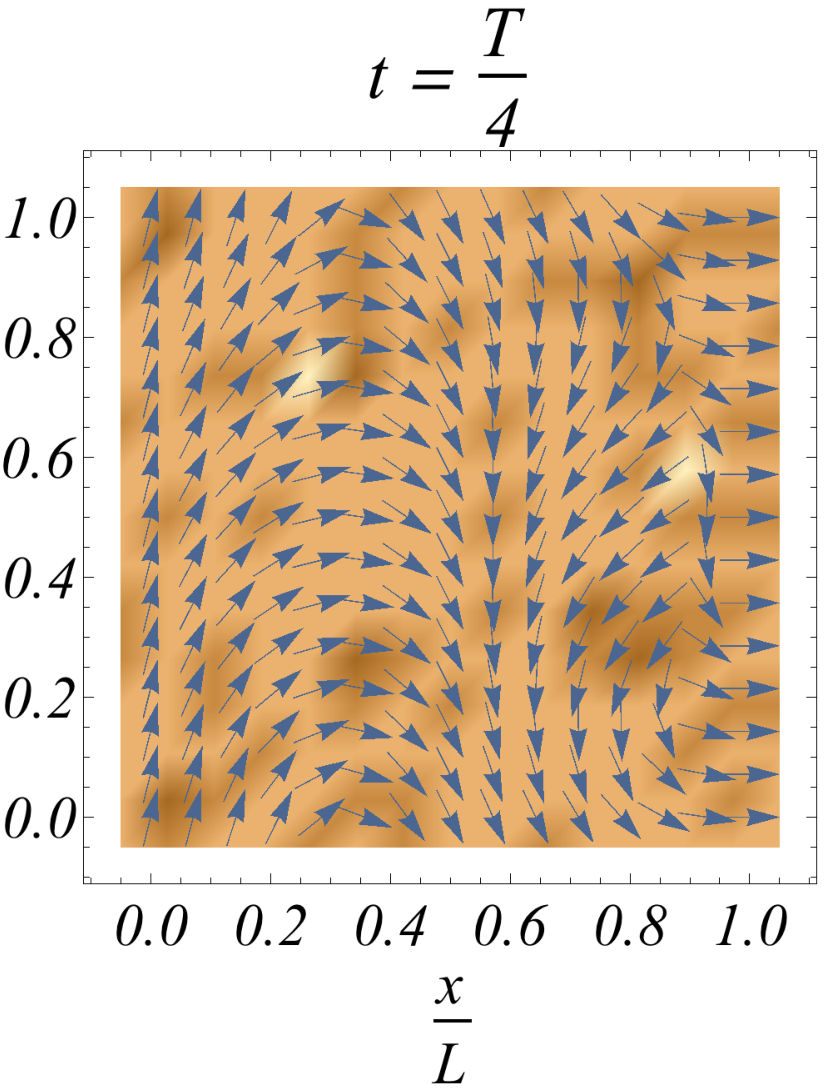}
	\includegraphics[width=0.28\linewidth]{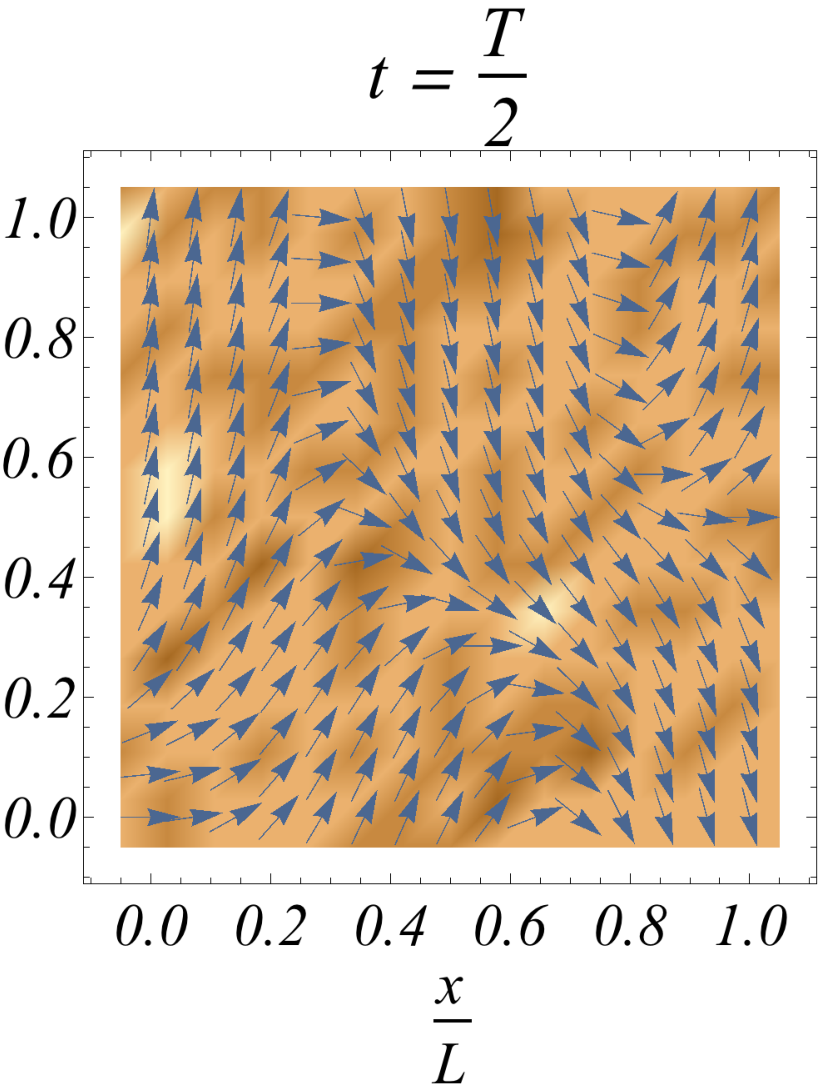}
	\caption{
	Director configurations of a deterministic typical trajectory of active nematic  ($\sqrt{A \over 2B}=1,{C' \over 2B}\sqrt{A\over 2B} =\frac12,\lambda=0$).
	(a) at $t=0$; (b) at $t=T/4$ and 
	(c) at $t=T/2$ where $T= 2 \pi / \nu$.
	\label{fig:director}
	}
\end{figure*}
Linear stability analysis shows that the uniform nematic state is unstable once , $ \exists \; \bfq$ s.t.  $Kq^2 + {\alpha' S_0 \cos 2 \theta_\bfq  N_\bfq \over \Delta_\bfq} < 0 $ 
where $N_\bfq=\left(1 + {\lambda \over 2} \cos 2 \theta_\bfq \right)$, $\Delta_\bfq= \eta  \left( 1 + {2 S_0^2 \over  \eta} +{\lambda^2 S_0^2 \over 2 \eta} + {2 \lambda S_0^2 \over \eta} \cos 2 \theta_\bfq \right)$, $\cos 2 \theta_\bfq = (q^2_2 -q_1^2)/q^2$,
$\alpha'=\alpha (1 + \lambda S_0)$, $q^2=|\bfq|^2$~\cite{AditiSimha2002,Voituriez2005}.   In what follows we restrict ourselves to $q<q_{max} \sim {\pi \over a}$, with $a$ a microscopic length, keeping the number of modes  finite. We restrict our analysis to  $\lambda <1$. For $\alpha >0$, the instability is driven by modes, $\bfq$ with $q_2 < q_1$ while for $\alpha  < 0$, it is driven by modes $\bfq$ with $q_2 > q_1$. At very small $|\alpha|$ only the lowest $q$ modes are unstable. 
To illustrate our approach,
we have  studied parameter ranges where  $|\alpha|$ small enough so that only a very small number of modes are linearly unstable: for $\alpha=\alpha_+>0$, such that modes  $\bfq_+=\frac{\pi}{L}(1,0), 2\bfq_+=\frac{\pi}{L}(2,0), \bfq_\uar=\frac{\pi}{L}(2,1)$ are unstable while for 
$\alpha=\alpha_-<0$, modes $\bfq_-=\frac{\pi}{L}(0,1),2\bfq_-=\frac{\pi}{L}(0,2), \bfq_\dar=\frac{\pi}{L}(1,2)$ are unstable.  
We treat each case, $\alpha_\pm$ separately.
\paragraph{Case $\alpha=\alpha_+$:} 
Analysis of the steady-state distribution and typical trajectories leads to the following observations.
All modes, $\tilde\bmQ(\bfq)$ apart from the unstable modes fluctuate about zero (they are equilibrium-like).  The unstable modes 
have amplitudes:  
$Q_1(\bfq_+) = 0,  Q_2(\bfq_+)\equiv  Q_{10}, \, Q_1(2\bfq_+) = 0,  Q_2(2\bfq_+)\equiv  Q_{20}, \, Q_1(\bfq_{\uar})\equiv  S_{21}, Q_2(\bfq_{\uar})\equiv  Q_{21}$, $\bmS_{21}=(S_{21},Q_{21})$ that on long timescales fluctuate about the deterministic trajectories with 
\bem
\dot Q_{10}=0 \, ; \, Q_{10}^2 \simeq  {C \over 3 B}  S_0 \, , \, \dot Q_{20}=0 \, ; \, Q_{20}^2 \simeq  {C \over 3 B}  S_0 \; , \; 
\nonumber\eem 
and 
\beq
\dt{} {\bmS_{21}}  = - \bfD(\bfq_{\uar}) \cdot \balpha (\bfq_{\uar}) \cdot \bmS_{21} \; ; \;  |\bmS_{21}|^2 \simeq  {C \over 3 B} S_0  \, ,
 \eeq
where $C= {\alpha' (1-\lambda/2) \over \eta} \ll A,B$ and keeping only leading order terms in $\alpha$. The mean amount of nematic order is renormalized by activity to  $\bar Q_1 = S=S_0+ \delta S_0   \simeq \sqrt{A\over 2B} - {C \over 2B}$. $\balpha,\bfD$ are matrices. They are  
\beq
\balpha(\bfq)=  \alpha' \left( \begin{array}{cc} 0 & -a \\ a & 0 
 \end{array}\right) \; , \; \bfD (\bfq) = \left( \begin{array}{cc}  D_{11} & D_{12} \\  D_{21} & D_{22}
 \end{array}\right) \; . \nonumber  
\eeq
where
 $D_{11}=\left( 1+ {\lambda^2 S_0^2 \over 2 \eta} \sin^2 2\theta_\bfq\right)$, $D_{22}=\left(1 + {2 S_0^2 \over \eta} N_\bfq^2 \right)$,   $D_{12}=D_{21}=-{\lambda S_0^2 \over \eta} \sin 2 \theta_\bfq N_\bfq$, $a= - {S_0 \sin 2 \theta_\bfq \over 2 \Delta_\bfq}$, $\sin 2 \theta_\bfq = 2 q_1 q_2/q^2$.
$\balpha$ is antisymmetric, hence any  term proportional to $\balpha$ {\em cannot} be written as the derivative of a scalar function and makes the system non-equilibrium as defined above.  $\bfD$ is a mobility matrix~\cite{Appendix}.  This leads to oscillatory behaviour of  mode $\bfq_\uar$ with frequency $\nu=|{\alpha' a}|\sqrt{\mbox{det}\left(\bfD(\bfq_\uar)\right)}$. Hence we can construct the evolution of the average dynamics of the nematic director as illustrated in Figure 2. When $A \simeq 0$,  anomalous fluctuations expected near critical points mean that these results must be augmented by RG analysis. 
As such points are rare, one expects to find few experiments in their vicinity~\cite{Sanchez2012}.

\paragraph{Case $\alpha=\alpha_-$:}  
Here the unstable modes
have amplitudes:  
$Q_1(\bfq_-) = 0,  Q_2(\bfq_-)\equiv Q_{01}, \; Q_1(2\bfq_-) = 0,  Q_2(2\bfq_-)\equiv Q_{02}, \;Q_1(\bfq_{\dar})\equiv  S_{12}, Q_2(\bfq_{\dar})\equiv   Q_{12}$, $\bmS_{12}=(S_{12},Q_{12})$.  As above these modes 
fluctuate about deterministic trajectories with $Q_{01},Q_{02},\bmS_{12}$ following the same equation as $Q_{10},Q_{20},\bmS_{21}$ respectively with $C$ replaced by $C'$ 
\section{Discussion}

\tl{Thermodynamic equilibrium is characterised by the macroscopic quantities of a system being stationary in time. Equilibrium statistical mechanics provides a link to microscopic degrees of freedom via a steady-state probability distribution of microstates (that maximises entropy). The utility of equilibrium statistical mechanics rests on the ability to express macroscopic quantities in terms of sums over microstates weighted by this probability distribution. 
In general, these sums are very difficult to evaluate, however their very existence justifies numerous approximations that can be made which allow many of these quantities to be 
calculated to a  controllable  accuracy using a variety of analytical and numerical techniques~\cite{ChaikinLubensky95}. 
}

\tl{
In this article we \tbl{show how}  to place a  \tbl{a number} 
of classical non-equilibrium systems on a similar footing. We find that in order to do this one must add a new dynamical aspect to 
the concept of the steady state. These non-equilibrium steady states are intrinsically dynamic in the sense that they are steady only at the level of the probability density of microstates. The probability density of states can only remain steady if the systems moves through the microstates in a particular deterministic manner.  The {\em generalised} steady states are thus characterised by two related quantities, a probability density and a dynamical system. Unsurprisingly, one is not in general able to show that  generalised steady states exist for every non-equilibrium system, however we are able to show when they can be found, under what conditions such generalised steady states are stable. Furthermore, 
we show that if the steady state distribution satisfies specific properties, then the system relaxes exponentially fast to that generalised steady state on a timescale that we can calculate. 
\tbl{This has been done 
by reformulating and extending 
some results from the mathematics of stochastic systems~\cite{Bakry1984,Bakry1994book,Markowich2000,Guionnet2002}. The presentation has been kept non-technical and physically intuitive. 
We hope this will lead to wide application in studies  of  realistic driven fluctuating physical systems, thus allowing the implications of experimental measurements on active matter systems to be more precisely quantified~\cite{Schaller2010,Palacci2013,Bricard2013}.}
In practice, one will be restricted to situations where one can only obtain approximate expressions for steady-state distributions and dynamical systems that characterise the steady states. The nonlinearity of the problem means that there will in general be more than one candidate steady state.  Hence one can use this relaxation timescale as a criterion to rank
steady states.}

\tl{Once in a generalised steady state, macroscopic quantities can like equilibrium systems be calculated as sums over microstates weighted by the steady state distribution. However unlike equilibrium systems in which they do not change in time, these macroscopic properties evolve in time according to the typical dynamical system. A trivial example of a typical dynamical system is one that is constant in time giving equilibrium-like behaviour. The simplest non-trivial example of a typical dynamical system which cannot be mapped to an equivalent equilibrium system is one  which shows cyclic motion. We illustrate our approach with a few examples of this type. In particular we obtain a new way to  characterise the behaviour of active nematics beyond the generic instability of active liquid crystals. We find that beyond the instability, some of the soft goldstone modes of the nematic, instead of fluctuating about zero become excited by activity and their amplitudes develop oscillatory behaviour. However the period of oscillation of each mode (which we explicitly calculate) is different. This leads to a highly dynamic but deterministic disordered structure of director orientations and fluid velocity on average which we can explicitly describe and predict.}

In a generalised steady state,
 the average long time behaviour  of 
 this class of non-equilibrium systems can thus be quantified by (1) picking an ensemble of initial conditions randomly from the steady state distribution $\rho$,  (2) following each realisation's evolution along the typical trajectory which goes through its initial point; (3) finally one can average over typical trajectories to obtain temporal correlations. 

We have studied dynamics in the overdamped limit in which momentum degrees of freedom are assumed to have relaxed to their steady-state values, however our analysis can also be extended to timescales for which momentum degrees of freedom are still relevant~\cite{TBL_unpub}. 
We conclude by noting that while we have found conditions in the form of  strict bounds on the derivatives of the steady state distribution, it is possible (and indeed expected) that with more sophisticated analysis, one can find more accurate bounds which will make these conditions valid for an even wider class of steady states~\cite{TBL_unpub}.

\begin{acknowledgments}  
I acknowledge the support of The Royal Society.
\end{acknowledgments}

\bibliography{../../biblio,../../library,../../active,../../library-air}

\appendix

\onecolumngrid

\section{Hopf oscillator}

\tl{
The oscillator has $\ham= -\frac{A}{2} |\vx|^2 +\frac B4 (|\vx|^2)^2 $ and $\vec v = (v_1,v_2)=( \Omega x_2 , -\Omega x_1 )$ and the condition  that determines the stationary distribution, $ \ds \rho (\vx) =\frac{1}{Z} \exp \left[{- {\ham (\vx) \over \theta} - \epsilon (\vx) } \right]$ is 
\beq
 \sum_{i=1}^2   L_i (\epsilon) =0 \label{eq:ss-brus}
 \eeq
where $L_i (\epsilon) = \theta (\nabla_i \epsilon)^2  +
 v_i \nabla_i \epsilon + \nabla_i \epsilon \nabla_i \ham 
 - \theta \nabla_i^2 \epsilon  $.
We look for a power series expansion for $ \epsilon(\vx)
$ 
and keep terms up to a particular order, e.g.
\[
\epsilon = \frac12 \vx \cdot \bmM \cdot  \vx  + \frac14 g |\vx|^4 \ldots \quad , \quad \bmM = \left(  \begin{array}{cc} m_1 & m_2 \\ m_4 & m_3 \end{array} \right) \, ,
\]
This can be substituted into eqn.~(\ref{eq:ss-cond}) to obtain a power series which can be set to zero term by term starting with the lowest powers and stopping at the highest powers kept in the expansion for $\epsilon$ to obtain simultaneous nonlinear equations for the coefficients, $m_i,g$.
We obtain the equations
\beqa
g &=&  0 \quad ,  \\
\theta (m_1 + m_3)  & =& 0  \quad , \\
-\frac12 m_2\Omega  -\frac12 m_4 \Omega - A m_1 + \theta \left ( \frac14 \left(m_2 + m_4 \right)^2 + m_1^2 \right) &=& 0  \quad , \\
\frac12 m_2 \Omega +\frac12 m_4 \Omega - A m_3 + \theta \left ( \frac14 \left(m_2 + m_4 \right)^2 + m_3^2 \right) &=& 0  \quad , \\
\Omega ( m_3 - m_1) +  A \left( m_2 + m_4 \right) -  \theta (m_2+m_4)  (m_1 + m_3) &=& 0  \quad , 
\eeqa
\tbl{giving us 5 equations for 5 unknowns. Solving the equations give
the solution described in the main text. The solution has $\caC (\vx) =0$.
}}

\section{Brusselator}

The Brusselator has $\ham=0$ and $\vec v = (v_x,v_y)=( \mu+x^2y - (\lambda+1)x, \lambda x -x^2y )$ \cite{Strogatz94} and hence the condition  that determines the stationary distribution, $\rho (x,y) =\frac{1}{Z}e^{-h(x,y)}$ is 
\beq
 \sum_{i=1}^2   L_i (h) =0 \label{eq:ss-brus}
 \eeq
where $L_i (h) = \theta (\nabla_i h)^2 + \nabla_i^2 \ham +
 \nabla_i h \left( 
 v_i-\nabla_i \ham  \right)- \theta \nabla_i^2 h   - \nabla_i v_i  $.
\tl{We look for a power series expansion for $h(x,y)
$ 
and keep terms up to a particular order, e.g.
\[
h = a_1 x + \frac{1}{2}  a_2 x^2 + \frac{1}{3}  a_3 x^3 + \frac14 a_4 x^4 + y \left( b_0  + b_1 x + \frac{1}{2}  b_2 x^2  + \frac13 b_3 x^3 \right) + \frac{y^2}{2} \left( c_0   + c_1  x   +  \frac12 c_2 x^2 \right) + \frac {y^3}{3} \left( d_0  + d_1 x \right) + \frac14 e_0 y^4 + \ldots
\]
}

\tl{
This can be substituted into eqn.~(\ref{eq:ss-cond}) to obtain a power series which can be set to zero term by term starting with the lowest powers to obtain simultaneous nonlinear equations for the coefficients, $a_i,b_i,c_i,d_i,e_i$.
We obtain the equations
\beqa
-1 - \lambda - a_1 \mu - a_1^2 \theta + a_2 \theta - b_0^2 \theta + 
   c_0 \theta &=&  0 
   \\  
 a_1 + a_1 \lambda - b_0 \lambda - a_2 \mu - 2 a_1 a_2 \theta + 2 a_3 \theta - 
   2 b_0 b_1 \theta + c_1 \theta &=&  0  
   \\ 
  - b_1 \mu - 2 a_1 b_1 \theta + b_2 \theta - 2 b_0 c_0 \theta + 2 d_0 \theta &=& 
  0 
  \\ 
  - 1 + a_2 + a_2 \lambda - b_1 \lambda - a_3 \mu - a_2^2 \theta - 
   2 a_1 a_3 \theta + 3 a_4 \theta - b_1^2 \theta - b_0 b_2 \theta + (c_2 \theta)/
   2 &=& 0  
   \\  
   - c_1 \mu - 2 b_1^2 \theta - 2 c_0^2 \theta - 2 a_1 c_1 \theta + 
     c_2 \theta - 4 b_0 d_0 \theta + 6 e_0 \theta   &= &  0  
     \\  
 2 + b_1 + b_1 \lambda - c_0 \lambda - b_2 \mu - 2 a_2 b_1 \theta - 
   2 a_1 b_2 \theta + 2 b_3 \theta - 2 b_1 c_0 \theta - 2 b_0 c_1 \theta + 
   2 d_1 \theta  & = &  0 
   \\  
 - a_1 + b_0 + b_2 + b_2 \lambda - c_1 \lambda - b_3 \mu - 2 a_3 b_1 \theta - 
   2 a_2 b_2 \theta - 2 a_1 b_3 \theta - b_2 c_0 \theta - 2 b_1 c_1 \theta - 
   b_0 c_2 \theta & = &  0 
   \\ 
 c_1 + c_1 \lambda - 2 d_0 \lambda - c_2 \mu - 4 b_1 b_2 \theta - 
     2 a_2 c_1 \theta - 4 c_0 c_1 \theta - 2 a_1 c_2 \theta - 4 b_1 d_0 \theta - 
     4 b_0 d_1 \theta  & = & 0  
     \\ 
 a_3 + a_3 \lambda -  (b_2 \lambda)/2 - a_4 \mu - 2 a_2 a_3 \theta - 
   2 a_1 a_4 \theta - b_1 b_2 \theta - (2 b_0 b_3 \theta)/3 & = &  0 
   \\  
 -  (d_1 \mu)/3  - b_1 c_1 \theta - 2 c_0 d_0 \theta - (2 a_1 d_1 \theta)/3 - 
   2 b_0 e_0 \theta & = &  0  
   \\ 
 a_4 + a_4 \lambda - (b_3 \lambda)/3 - a_3^2 \theta - 2 a_2 a_4 \theta - ( b_2^2 \theta)/4 - (2 b_1 b_3 \theta)/3 & = &  0  
   \\  
 - (c_1^2 \theta)/4  - d_0^2 \theta - (2 b_1 d_1 \theta)/3 - 
   2 c_0 e_0 \theta  & = &  0 
   \\  
 - a_2 + b_1 + b_3 + b_3 \lambda - (c_2 \lambda)/2 - 2 a_4 b_1 \theta - 
   2 a_3 b_2 \theta - 2 a_2 b_3 \theta - (2 b_3 c_0 \theta)/3 - b_2 c_1 \theta - 
   b_1 c_2 \theta & = &  0 
   \\  
  -2 b_1 + 2 c_0 + c_2 + c_2 \lambda - 2 d_1 \lambda - 2 b_2^2 \theta - 
     4 b_1 b_3 \theta - 2 a_3 c_1 \theta - 2 c_1^2 \theta - 2 a_2 c_2 \theta - 
     2 c_0 c_2 \theta - 2 b_2 d_0 \theta - 4 b_1 d_1 \theta   & = &  0 \quad . 
\eeqa
These equations may be solved for  the specific  parameters  considered in the main text.  These nonlinear equations yield a large number of possible solutions for the coupling constants, $\{ a_i,b_i,c_i,d_i,e_i \}$, all having $\caC (x,y) \ne 0$. For  parameters $\mu=1,\lambda=3,\theta= \frac12 (0.1)^2$, there are 90 possible solutions for sets of constants, $\{ a_i,b_i,c_i,d_i,e_i \}$, and for  parameters $\mu=1,\lambda=3,\theta= \frac12 (0.2)^2$, there are 96 possible sets of constants. For $\theta$ small, we find one family of solutions that have $\caC > 0$ in the region close to the attractor (a limit cycle), i.e. they correspond to a stable non-equilibrium steady state, however as $\theta$ increases that is no longer the case. 
For example, when $\mu=1,\lambda=3,\theta= \frac12 (0.2)^2$, we obtain $ a_1 =  -5.99184, a_2 =  0.673657, a_3 =  0.153112, a_4 =  -0.0245148, b_0 
=  -8.10039, b_1 =  0.541483, b_2 =  0.407974, b_3 =  -0.0979923, c_0 =  1.25294, c_1 =  -0.15968, c_2 =  -0.344532, d_0 =  -0.0606681, d_1 =  0.011561, e_0 =  -0.00567803 $, while  for $\mu=1,\lambda=3,\theta=\frac12 (0.1)^2$, we find $a_1 =  -4.25216, a_2 =  0.309472, a_3 =  0.137682, a_4 =  -0.00944351, 
 b_0 =  -5.76712, b_1 =  0.038488, b_2 =  0.372361, b_3 =  -0.038008, 
 c_0 =  0.598897, c_1 =  0.000139246, c_2 =  -0.282185, d_0 =  0.0450479, 
 d_1 =  -0.0011471, e_0 =  -0.00166964$. For $\mu=1,\lambda=3,\theta= \frac12 (0.4)^2$, we can no longer find a stable generalised steady state.
%
Once we have the parameters $\{ a_i,b_i,c_i,d_i,e_i \}$, 
we can construct approximate expressions for $h$, the steady-state density and the typical trajectories. 
These can be improved by including more terms in the expansion for $h$.}

\section{d=2 active nematic}

The equations of motion for the  fields $\bmv (\bfr,t),\bmQ(\bfr,t)$, the fluid velocity and the traceless symmetric nematic order parameter respectively, are those of  incompressible viscous nematodynamics  augmented to include activity~\cite{DeGennesProst93}.
%
The fields are explicitly
\beq
 \bm{Q} =  \left( \begin{array}{cc} Q_{11}  & Q_{12}  \\
  Q_{21} & Q_{22} \end{array} \right) 
 =  \left( \begin{array}{cc} Q_1 (\bfr,t) & Q_2 (\bfr,t) \\
  Q_2 (\bfr,t) & - Q_1 (\bfr,t)
\end{array} \right) \quad , \quad \bm{v} = \left( v_1 (\bfr,t), v_2 (\bfr,t) \right) \; .
\eeq
Using the velocity field we can define the symmetric and asymmetric parts of the strain rate tensor: $u_{ij}=\frac12 ( \partial_i v_j +  \partial_j v_i )$, $\omega_{ij}=\frac12 ( \partial_i v_j -  \partial_j v_i)$ respectively.
The dynamics can be reduced to coupled equations for $Q_1(\bfr,t)=Q_{11}$, $Q_2(\bfr,t)=Q_{12}$, $u_1(\bfr,t)=u_{11}$, $u_2(\bfr,t)=u_{12}$ and $\omega (\bfr,t)=\omega_{12}$. We consider the system in a square box of area ${\cal A}=L^2$, $\bfr=(x,y), \{0 \le x,y \le L\}$ with $Q_1=\bar Q_1=S_0$ and $Q_2=\bar Q_2=0$ on the boundary $\partial {\cal A}$ where $S_0=\sqrt{2A/B}$. We consider 
deviations of $Q_i=\bar Q_i + \delta Q_i$ around the value on the boundary and $u_1,u_2,\omega$ around a stationary fluid. We have boundary conditions, $\delta Q_1=0, 
\nvec \cdot \nabla \delta Q_2 =0,
u_1=0,\nvec \cdot \nabla u_2=0, \nvec \cdot \nabla \omega=0$ on $\partial {\cal A}$ which are also respected by the fluctuations ($\nvec$ is local normal to $\partial {\cal A}$). Both $u_1,u_2$ can be expressed in terms of $\omega$: $u_2= (\partial_2^2 - \partial_1^2)\nabla^{-2} \omega, u_1= 2 \partial_1\partial_2 \nabla^{-2} \omega$ so there is only one independent component of the strain rate tensor.
%
The linearity of the Stokes eqn. (\ref{eq:stokes}) means that the vorticity is slaved to director
\beqa
\omega (\bfr,t)  && = {\nabla^{-2} \over \eta} \left[ \partial_1 \partial_2  \left( \tilde\alpha + \lambda S_0 \left(A- {B\norm{\bmQ}^2 / 2} - K \nabla^2 \right)Q_1  \right) \right. \nonumber \\ && + \frac12 (\partial_2^2-\partial_1^2) \left(\alpha' + \lambda S_0  \left(A- {B\norm{\bmQ}^2 /  2} - K \nabla^2\right) Q_2 \right)  \nonumber \\ &&   \left .  + S_0 K (\nabla^2)^2  Q_2  +  \frac12 \left( \partial_2 \tilde \xi_1^v -  \partial_1 \tilde \xi_2^v \right) \right]
\eeqa
where $\tilde \alpha = \alpha'+ 4 \lambda S_0^3 B$,  $\alpha'=\alpha (1 + \lambda S_0)$.

This means that the only truly independent fields are the $\tilde Q_1,\tilde Q_2$.
Taking Fourier transforms,
\beqa
 \tilde Q_1 (\bfq,t) =  && 
 \int_\bfr \Psi_\bfq(\bfr) \delta Q_1 (\bfr,t) \, , \nonumber  \\
\;  \tilde Q_2 (\bfq,t)  = && 
 \int_\bfr \Phi_\bfq(\bfr) \delta Q_2 (\bfr,t) \, , \nonumber \\
\; \tilde \omega (\bfq,t) = && 
\int_\bfr \Phi_\bfq(\bfr) \omega (\bfr,t) \, , \; \nonumber \\
\tilde \xi^Q_1 (\bfq,t) =  && 
 \int_\bfr \Psi_\bfq(\bfr)  \xi^Q_1 (\bfr,t) \, , \nonumber  \\
\;  \tilde \xi^Q_2 (\bfq,t)  = && 
 \int_\bfr \Phi_\bfq(\bfr)  \xi^Q_2 (\bfr,t) \, , \nonumber \\
\;  \tilde \xi^v_i (\bfq,t)  = && 
 \int_\bfr {1 \over q_i}\partial_i\Psi_\bfq(\bfr)  \xi^v_i (\bfr,t) \, , \nonumber \eeqa
%
 where  $\bfq=(q_1,q_2)=\frac{\pi}{L}(n,m)$ with $n,m \in {\Bbb Z}^+$, $\Psi_\bfq(\bfr)={\cal N} \sin (q_1 x) \sin (q_2 y) $, $\Phi_\bfq(\bfr)= {\cal N}\cos (q_1 x) \cos (q_2 y)$ with $\cal N$ chosen so that $ \frac{1}{L^2} \int_\bfr \Psi_\bfq^2 =  \frac{1}{L^2}  \int_\bfr \Phi_\bfq^2 =1$. 
 
 Linear stability analysis shows that the uniform nematic state is unstable once , $ \exists \; \bfq$ s.t.  $Kq^2 + {\alpha' S_0 \cos 2 \theta_\bfq  N_\bfq \over \Delta_\bfq} < 0 $ 
where $N_\bfq=\left(1 + {\lambda \over 2} \cos 2 \theta_\bfq \right)$, $\Delta_\bfq= \eta  \left( 1 + {2 S_0^2 \over  \eta} +{\lambda^2 S_0^2 \over 2 \eta} + {2 \lambda S_0^2 \over \eta} \cos 2 \theta_\bfq \right)$, $\cos 2 \theta_\bfq = (q^2_2 -q_1^2)/q^2$.

We now consider the case when $\alpha=\alpha_+>0$, such that modes with $\bfq_+=\frac{\pi}{L}(1,0), 2\bfq_+=\frac{\pi}{L}(2,0), \bfq_\uar=\frac{\pi}{L}(2,1)$ are linearly unstable. 
The unstable modes
have amplitudes:  
$Q_1(\bfq_+) = 0 , \;   Q_2(\bfq_+)\equiv  Q_{10}, \; Q_1(2\bfq_+) = 0, \;  Q_2(2\bfq_+)\equiv  Q_{20}, \; Q_1(\bfq_{\uar})\equiv \bar S_{21}, \; Q_2(\bfq_{\uar})\equiv \bar Q_{21}$, $\bmS_{21}=(S_{21},Q_{21})$. 
In addition we must consider the average amount of nematic order $\bar Q_1$. We define $\tilde\bmQ(\bfq)=(\tilde Q_1 (\bfq), \tilde Q_2 (\bfq))$.

We get equations for the 
modes
\beq
\dt {}{\tilde \bmQ(\bfq)}  =  \bfD(\bfq) \cdot \left[ -{ \partial \ham \over \partial \tilde \bmQ (\bfq)} - \balpha (\bfq) \cdot \tilde \bmQ (\bfq) \right] + \bmW(\bfq,t) \, ,
\eeq
with noise  $\bmW =\left( W_1 , W_2 \right) $, and $\ham(\bmQ,\alpha)$ is an effective non-equilibrium "Hamiltonian". The matrix $\balpha$ is antisymmetric and hence the term proportional to $\balpha$ cannot be written as the derivative of a scalar function and makes the system non-equilibrium in the manner defined in the main text.
\beq
\balpha(\bfq)=  \alpha' \left( \begin{array}{cc} 0 & -a \\ a & 0 
 \end{array}\right) \; , \; \bfD (\bfq) = \left( \begin{array}{cc}  D_{11} & D_{12} \\  D_{21} & D_{22}
 \end{array}\right) \; .
\eeq
The components of the mobility matrix are
 $D_{11}=\left( 1+ {\lambda^2 S_0^2 \over 2 \eta} \sin^2 2\theta_\bfq\right)$, $D_{22}=\left(1 + {2 S_0^2 \over \eta} N_\bfq^2 \right)$,   $D_{12}=D_{21}=-{\lambda S_0^2 \over \eta} \sin 2 \theta_\bfq N_\bfq$, $a= - {S_0 \sin 2 \theta_\bfq \over 2 \Delta_\bfq}$ where $\sin 2 \theta_\bfq = 2 q_1 q_2/q^2$,
and the fluctuations have moments
\beq
\aver{\bmW} =0 \quad , \quad  \; \aver{W_i (\bfq,t)W_j (\bfk,t)}=2 
 D_{ij} \delta_{\bfq\bfk} \delta(t-t') \; .
\eeq
Defining $\norm{\bmS}^2 = \bar Q_1^2 + Q_{10}^2  + Q_{20}^2 + \norm{\bmS_{21}}^2$, 
the non-equilibrium Hamiltonian, $\ham[\tilde \bmQ,\alpha]$ is  approximated as
\beqa
{\ham \over L^2} & \simeq & -\frac{A}{2} \norm{\bmS}^2 + \frac{B}{4} \left[ \left( \norm{\bmS}^2 \right)^2 + \frac12 Q_{10}^4 + \frac14 \left( Q_{21}^4 + S_{21}^4\right)\right] + \frac12  \bmS_{21} \cdot \bmk(\bfq_\uar) \cdot \bmS_{21} \nonumber \\ && 
+ \frac12  \left( K q_+^2 + \alpha' k_{22}(\bfq_+)\right) Q_{10}^2  + \frac12  \left( 4 K q_+^2 + \alpha' k_{22}(2\bfq_+)\right) Q_{20}^2  + \frac12\sum_{\bfq '}^{q_{max}} \tilde\bmQ(\bfq') \cdot \bmk(\bfq') \cdot \tilde \bmQ(\bfq') \nonumber \\ &&
\eeqa
where the sum $\ds \sum_{\bfq'}^{q_{max}}$ is over all the stable modes and 
\beq
\bmk (\bfq)= \left( \begin{array}{cc} K q^2 + 4B S_0^2 & 0 \\ 0 & K q^2 
 \end{array}\right) + \alpha'   \left( \begin{array}{cc} k_{11} (\bfq)  & k_{12}  (\bfq)   \\ k_{21}  (\bfq)    & k_{22}   (\bfq) 
 \end{array}\right) \; ,
\eeq 
with $k_{11}={S_0 \lambda \sin^2 2 \theta_\bfq  \over 2 \Delta_\bfq}$, $k_{22}={S_0 \cos 2 \theta_\bfq N_\bfq \over \Delta_\bfq}$,  $k_{12}=k_{21}= - {S_0 \sin 2 \theta_\bfq N_\bfq \over 2 \Delta_\bfq}$. It is noteworthy that for the stable modes both eigenvalues of $\bmk(\bfq)$ are positive while for the unstable modes, one or more is negative. 
\tl{
Hence the ``non-equilibrium effective Hamiltonian" can be written as $\ham = \sum_\bfq \ham_\bfq$ where $\ham_\bfq = -  \frac12 A_\bfq |\tilde \bmQ(\bfq)|^2  +  \frac14 B_\bfq | \tilde \bmQ(\bfq)|^4 $ where $A_\bfq >0$ for the unstable modes and $A_\bfq < 0$ for the stable modes, ($B_\bfq>0$ for all modes). This means that any pair of unstable modes $\tilde \bmQ (\bfq)$ can be described as effective Hopf oscillators.}

The 
"Hamiltonian" , $\ham$ is minimised for a manifold of mode amplitudes given by  
\beq
\quad Q_{10}^2=  f_{10}{C \over 2B} S_0  \, , \,  Q_{20}^2=  f_{20}{C \over 2B} S_0  \, , \, |\bmS_{21}|^2=  f_{21}{C \over 2B}S_0 \, , \, \bar Q_1 = S_0 - \frac12 \left(f_{21} + f_{10} + f_{20} \right){C \over 2B} \, , \,  \tilde Q(\bfq') = 0 \, .
\eeq
where we have taken $C \ll A,B$ and ignore terms of $O(C^2)$, and taken the limit of large system size, $L \gg 1$. The values of the scaling factors, $f_{ij}$ are functions of all the parameters in a manner dependent on the closure approximations taken in finding the minimum. In the main text we have taken all $f_{ij}=2/3$. Hence we consider the dynamics in the vicinity of this minimum and obtain an expression for $h$.
Our method for obtaining an approximate expression for $h= \ham + \epsilon$ is as follows. We write down a Taylor series expansion
expansion for $\epsilon[ \bmQ,\alpha ]$ keeping terms up to linear order in $\alpha$ and quadratic order in the fields, $\tilde \bmQ(\bfq)$. Truncating the expansion at low order is a reasonable approximation as long as $\bmQ$ is not too large. Clearly such a method can systematically improved by including higher order terms.
We thus look for an expression for $\epsilon$ of the form 
\beq
\epsilon = \sum_{\bfq}^{q_{max}} \frac12  \bmQ (\bfq) \cdot \bm {\bmM}(\bfq,\alpha)  \cdot  \bmQ (\bfq)  + \cdots \; ,
\eeq
which satisfies the 
condition for the stationary probability density $ \ds \rho[\bmQ] = \frac{1}{Z} \exp \left(- h [ \bmQ] \right)$ :
\beq
 \sum_{\bfq}^{q_{max}}   \Tr \left( \bfD(\bfq) \cdot {\bf L} (\bfq) \right) =0 \label{eq:ss-cond2}
 \eeq
where 
\beq
{\bf L} (\bfq)= \theta \left ( {\partial h \over \partial \tilde \bmQ (\bfq) }\right)^2 + {\partial^2 \ham \over \partial \tilde \bmQ (\bfq)^2 } +
{\partial h \over \partial \tilde \bmQ (\bfq) }  \left( 
\balpha \cdot \tilde \bmQ - {\partial \ham \over \partial \tilde \bmQ (\bfq) }  \right)-  \theta  {\partial^2 h \over \partial \tilde \bmQ (\bfq)^2 }    - \balpha(\bfq) \; \;  .
 \eeq
To satisfy the stationarity condition, $\bfD \cdot \bmM$ must also be traceless and symmetric. 
\tl{A matrix, $\bmM$ can always be found that keeps the minimum of $\ham$ unchanged so all modes apart from the linearly unstable ones fluctuate about zero. The solution for $h$ corresponding to this matrix $\bmM$ has  $\caC  =0$.}

Therefore all modes except $\bfq_+,2 \bfq_+, \bfq_\uar$ fluctuate about zero and make no contribution to the typical trajectories of the system. Hence we obtain the deterministic equations in the main text for the typical trajectories for these  modes.
A similar analysis can be performed for the $\alpha_-$ case.

\end{document}